\documentclass[conference]{IEEEtran}
\IEEEoverridecommandlockouts
\usepackage{amsmath,amssymb,amsfonts}

\usepackage{tabularx}
\usepackage{algorithmic}
\usepackage{graphicx}
\usepackage{textcomp}
\usepackage{todonotes}

\usepackage{enumitem}

\usepackage{verbatimbox}
\usepackage{multirow}
\usepackage{verbatim}
\usepackage{fancyvrb}
\usepackage{listings}

\usepackage{pgf-pie}
\usepackage{jigsaw}

\usepackage{xcolor}
\usepackage{booktabs}
\usepackage{makecell}
\usepackage{siunitx}
\usepackage[dvipsnames]{xcolor}
\usepackage{listings}
\usepackage{tikz}

\usetikzlibrary{positioning, arrows.meta, calc, shapes, decorations.pathreplacing}

\usepackage[sorting = none, style = numeric-comp, backend = biber, style=ieee]{biblatex}
\newcommand{\drawkey}[3]{
    \draw[line width=0.07cm, draw=#2] (#1) circle [radius=0.15cm];
    \draw[line width=0.07cm, draw=#2] (#1 -0.15) -- ++(0,-0.4);
    \draw[line width=0.07cm, draw=#2] (#1 -0.35) -- ++(-0.2,0);
    \draw[line width=0.07cm, draw=#2] (#1 -0.51) -- ++(-0.2,0);
    \node[below, text=#2] at (#1 -0.6) {#3}
}

\usepackage{cooltooltips}
\usepackage{graphicx}
\usepackage{color}
\usepackage{hyperref}

\usepackage{pifont}% http://ctan.org/pkg/pifont
\newcommand{\cmark}{{\ding{52}}}%
\newcommand{\xmark}{{\ding{55}}}%

\begin{document}

\title{ Implementation and transition to post-quantum cryptography of the Minimal IKE protocol
\thanks{
This work was partially supported by project SERICS (PE00000014) under the MUR National Recovery and Resilience Plan, funded by the European Union - Next Generation EU and by MUR under the Italian Fund for Applied Science (FISA 2022), Call for tender No. 1405 published on 13-09-2022 - project title “Quantum-safe cryptographic tools for the protection of national data and information technology assets” (QSAFEIT) - No. FISA 2022-00618 (CUP I33C24000520001), Grant Assignment Decree no. 15461 adopted on 02.08.2024.}
}

\author{\IEEEauthorblockN{Davide De Zuane\IEEEauthorrefmark{1}, Paolo Santini\IEEEauthorrefmark{2}, Marco Baldi\IEEEauthorrefmark{2}}
\IEEEauthorblockA{\IEEEauthorrefmark{1}
\textit{IMT School for Advanced Studies},
Lucca, Italy}
\IEEEauthorblockA{\IEEEauthorrefmark{2}
\textit{Università Politecnica delle Marche},
Ancona, Italy\\
e-mail: davide.dezuane@imtlucca.it, \{p.santini, m.baldi\}@univpm.it}
}

\maketitle

\begin{abstract}
This paper concerns the Minimal Internet Key Exchange (IKE) protocol, which has received little attention to date, despite its potential to make the best-known IKE protocol sufficiently lightweight to be also applied in contexts where it is currently prohibitive, due to its large footprint.
First, we introduce and describe Colibri, an efficient, open-source implementation of the Minimal IKE protocol, which allows us to quantitatively assess its real advantages in terms of lightness.
Then we introduce a post-quantum variant of the Minimal IKE protocol, which is essential to make it contemporary, and assess it through Colibri.
We demonstrate that the protocol performance remains excellent even in such a more challenging context, making it suitable for deploying pervasive and quantum-resistant virtual private networks.
\end{abstract}


\section{Introduction}

Internet Protocol Security (IPsec) is a family of protocols that introduces security techniques at layer 3 of the network protocol stack. Thanks to IPsec, communications achieving confidentiality, integrity and authentication can be instantiated, thus protecting IP communications from attacks such as eavesdropping and tampering \cite{NISTIPsec}. 
Formally, IPsec protocols are used to create Security Associations (SAs), which can be thought of as sets of parameters defining how data should be protected during communication between two entities of a network, thus transforming IP from a stateless protocol to a stateful protocol.
In this paper, we focus on the Internet Key Exchange (IKE) protocol, which is a fundamental protocol within IPsec, enabling the secure creation of SAs through negotiation of the SA parameters, as well as authentication and dynamic distribution of cryptographic keys \cite{Cremers2011}.

IPsec makes obviously a massive use of cryptography.
In particular, asymmetric cryptography is the only way to achieve the desired functionalities without any pre-shared secret. 
For instance, an IPsec protocol is used to make communicating endpoints agree on a shared secret from which keys for symmetric encryption are derived. This is made possible by the use of asymmetric cryptographic algorithms such as Diffie-Hellman (DH).
Analogously, public-key certificates and digital signatures are employed to authenticate the two endpoints.
\medskip

\paragraph*{Transitioning to post-quantum cryptography}

The cryptographic world is currently dealing with a new threat imposed by quantum computers.
While for symmetric encryption schemes and hash functions quantum algorithms do not pose a relevant threat, the situation is completely different for what concerns asymmetric algorithms.
Indeed, most of them can be broken efficiently using Shor's algorithm \cite{Shor}.
Post-quantum cryptography, which provides cryptographic algorithms that are resistant to both classical and quantum computers, is largely recognized as the most promising solution to face this threat.
As it is well known, the U.S. National Institute of Standards and Technology (NIST) started an international process of selecting and standardizing post-quantum cryptographic algorithms back in 2016 \cite{NISTPQC}.
On March 11, 2025, this process has officially been concluded, with five algorithms (two Key Encapsulation Mechanisms (KEMs) and three digital signatures) being declared as the first standards for post-quantum cryptography.
To render existing applications and devices quantum secure, one should ideally replace all asymmetric algorithms with their post-quantum counterparts.
This, however, poses some evident technical challenges and issues.
Indeed, post-quantum algorithms typically requires more bandwidth occupation (indeed, ciphertexts and public keys are larger).
Moreover, for some functionalities such as non-interactive key exchanges à la DH, we do not still have any post-quantum replacement.
\medskip

\paragraph*{Minimal IKE}

Following the above considerations, for some applications such as those involving IoT devices, the use of a post-quantum variant of IKE seems rather complicated.
A solution in this case may be that of relying on Minimal IKE \cite{rfc7815}, which is a protocol aimed at implementing a lightweight version of an IKE initiator.
Basically, the idea is that of dropping some of the functionalities which would be provided by a fully fledged implementation of IKE, to reduce both the bandwidth  occupation as well as the memory footprint.
Thus, Minimal IKE seems like a very promising solution to complete the transition to post-quantum cryptography in many cases.
However, at the best of our knowledge, currently there is no efficient, open source software implementation of Minimal IKE.
Perhaps because of this, studies that investigate the performance of Minimal IKE with post-quantum cryptosystems are also missing, to the best of our knowledge.

\subsection{Contribution and paper organization}

This paper fills the gaps we have identified above.
First, we present Colibri, the first fully fledged implementation of Minimal IKE.
We describe the implementation choices and optimizations that make Colibri very lightweight and suitable for execution even on devices with limited resources.
Using Colibri, we are able to propose the first post-quantum variant of Minimal IKE and analyze its performance in comparison with both its quantum-vulnerable version and the fully fledged version of IKEv2.
We provide public open-source code for both Colibri and the testing environment used for running our benchmarks\footnote{\url{https://github.com/secomms/colibri}}.

The paper is organized as follows:
Section \ref{sec:background} introduces the IKEv2 protocol.
Section \ref{sec:minimalIKE} describes the Minimal IKE protocol, while Section \ref{sec:PQCtransition} elaborates on its transition to post-quantum cryptography.
Section \ref{sec:implementation} describes the implementation of Colibri and assesses its performance both in the classical and post-quantum domains.
Finally, Section \ref{sec:conclusion} reports some conclusive remarks and anticipates some possible directions for future work.

\section{Internet Key Exchange v2}
\label{sec:background}

The IKEv2 protocol, specified in RFC 7296 \cite{rfc7296}, is an Authenticated Key Establishment (AKE) protocol that establishes cryptographic parameters and keying material for IPsec session. It operates through two main exchanges:

\begin{enumerate}[label=\Roman*.]
    \item \texttt{IKE\_SA\_INIT}: negotiates cryptographic algorithms and performs key establishment—typically via Diffie–Hellman—to derive a shared secret.
    \item \texttt{IKE\_AUTH}: provides mutual authentication and finalizes the creation of IPsec Security Associations (SAs), ensuring the shared key is bound to the authenticated identities. Ensuring that the shared key was established with the intended party.
\end{enumerate}

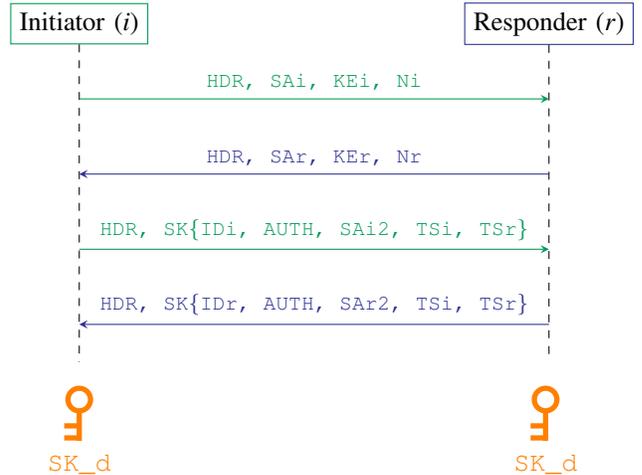
\begin{figure}[htbp!]
    \centering
    \begin{tikzpicture}[node distance=1.2cm]
        % Initiator and Responder
        \node (entity1) [draw=ForestGreen, rectangle] {Initiator (\textit{i})};
        \node (entity2) [draw=Blue, rectangle, right=of entity1, xshift=3cm] {Responder (\textit{r})};
        \draw[dashed] (entity1) -- ++(0,-4.5) coordinate (vertical1);
        \draw[dashed] (entity2) -- ++(0,-4.5);
        % IKE_INIT_SA
        \draw[-stealth, ForestGreen] (entity1) ++(0,-1) -- (entity2 |-,-1) node[midway, above, text=ForestGreen, font=\footnotesize] { \texttt{HDR, SAi, KEi, Ni} };
        \draw[stealth-, Blue] (entity1) ++(0,-2) -- (entity2 |-,-2) node[midway, above, text=Blue, font=\footnotesize] { \texttt{HDR, SAr, KEr, Nr} };
        % IKE_AUTH
        \draw[-stealth, ForestGreen] (entity1) ++(0,-3) -- (entity2 |-,-3) node[midway, above, text=ForestGreen, font=\footnotesize] { \texttt{HDR, SK\{IDi, AUTH, SAi2, TSi, TSr\}}};
        \draw[stealth-, Blue] (entity1) ++(0,-4) -- (entity2 |-,-4) node[midway, above, text=Blue, font=\footnotesize] {\texttt{HDR, SK\{IDr, AUTH, SAr2, TSi, TSr\}}};
        \drawkey{0,-5}{orange}{\texttt{SK\_d}};
        \drawkey{6.2,-5}{orange}{\texttt{SK\_d}};
    \end{tikzpicture}
    \caption{Main IKEv2 protocol exchanges}
    \label{fig:ike_exchange}
\end{figure}

\medskip
The simplest form of the protocol consists of the two main exchanges described above, each composed of an ordered sequence of payloads conveying specific information for initialization and authentication. Table~\ref{tab:ike_payloads} summarizes the most commonly used payload types and their functions. While the table presents each payload in a generic form, some payloads are configuration-dependent and independently generated by each peer. To indicate the origin of each payload, we adopt a subscript notation: payloads sent by the initiator are denoted with \texttt{i}, and those sent by the responder are denoted with \texttt{r}, as illustrated in Fig.~\ref{fig:ike_exchange} and in subsequent references.

\renewcommand{\arraystretch}{1.2}
\begin{table}[htbp]
    \centering    
    \caption{IKEv2 Payload Types}
    \label{tab:ike_payloads}
    \begin{tabular}{@{}p{1cm} p{6.8cm}@{}}
        \toprule
        \textbf{Name} & \textbf{Semantic} \\
        \midrule
        \texttt{HDR}      & Header of the message (not a payload). \\
        \texttt{SA}       & Security Association parameters (algorithms, DH groups).    \\
        \texttt{KE}       & Public key material for the key establishment protocol.      \\
        \texttt{N}        & Nonces selected by the initiator and responder. \\
        \texttt{ID}       & Identities to be used in authenticating the peer. \\
        \texttt{CERT}     & Public key certificates for authentication.                                 \\
        \texttt{AUTH}     & Payload that will be used to check the identity.                \\
        \texttt{TS}       & Traffic Selectors (adresses, ports).                           \\
        \texttt{SK\{..\}} & Symmetric authenticated encryption function.\\
        \bottomrule
    \end{tabular}
\end{table}
\medskip

IKEv2 is a highly flexible protocol supporting multiple methods for key establishment and peer authentication. Depending on the deployment scenario, it allows authentication via \textit{pre-shared keys} (PSK), \textit{digital signatures}, or the \textit{Extensible Authentication Protocol} (EAP). In addition to the two core exchanges, IKEv2 defines \texttt{CREATE\_CHILD\_SA}, used to establish additional Child SAs or rekey existing ones (including the IKE SA) for seamless key rotation, and \texttt{INFORMATIONAL}, which handles control messages such as SA deletion, error reporting, and other management notifications.

\section{Minimal IKE}
\label{sec:minimalIKE}

The \textit{Minimal IKE} variant, formalized in RFC 7815~\cite{rfc7815}, represents a streamlined subset of the IKEv2 protocol specifically designed for resource-constrained environments.

Over time, the IKEv2 protocol has gradually shifted from a compact negotiation mechanism into a comprehensive and extensible framework—powerful, but also more resource-intensive and harder to implement fully. 
With the increasing spread of IoT and embedded technologies, there is growing interest in securing IP communications in highly resource-constrained devices. In this context, IPsec remains a compelling candidate, as it is already integrated into most operating system kernels and can secure traffic at the network layer. Minimal IKE, by removing non-essential features of IKEv2, offers a lightweight negotiation layer that makes the use of IPsec feasible even in low-power or embedded platforms.
Next we briefly summarize the main features of Minimal IKE.

\medskip

\noindent
\textbf{Core Exchanges}: Minimal IKE processes only \texttt{INIT} and \texttt{AUTH}; all other exchange types are ignored or trigger empty/error responses, avoiding extra processing overhead.
\medskip

\noindent
\textbf{Feature Drop}: The Minimal IKE protocol omits optional extensions negotiated via notifications, reducing implementation surface and eliminating extra branching. 
A detailed comparison of the features included in the Minimal IKE protocol compared to the fully fledged IKEv2 protocol is reported in Table~\ref{tab:ikev2_vs_minimal}.
\medskip

\noindent
\textbf{Static Structure}: Payloads, algorithms, and payload ordering are fixed. This static design simplifies parsing, removes runtime negotiation, and lowers the risk of configuration errors and parsing vulnerabilities.
\medskip

\begin{table*}[htbp]
    \centering
    \renewcommand{\arraystretch}{1.3}
    \caption{Comparison between Full IKEv2 and Minimal IKE}
    \label{tab:ikev2_vs_minimal}
    \begin{tabular}{@{}p{4.2cm} p{5.5cm} p{5.5cm}@{}}
        \toprule
        \textbf{Feature} & \textbf{Full IKEv2 (RFC 7296)} & \textbf{Minimal IKE (RFC 7815)} \\
        \midrule
        Supported Exchanges & All & Only INIT and AUTH\\
        
        Notification Payloads & Fully supported for negotiating extensions & Ignored or minimally handled \\
        
        Cipher Suite & Dynamic negotiation of supported suites & Single static suite (predefined by configuration) \\
        
        Message Parsing & Full dynamic parsing of payloads and ordering & Static structure; fixed payload order and content \\
        
        Authentication Methods & PSK, Digital Signatures, EAP & PSK or minimal signature support \\
        
        Mobility Support (MOBIKE) & Fully supported (RFC 4555) & Not supported \\
        
        Fragmentation Support & Optional (RFC 7383) & Not supported \\
        
        Re-keying & Via \texttt{CREATE\_CHILD\_SA} & Not supported \\
        
        Target Environment & General-purpose systems (e.g., VPNs, firewalls) & Constrained devices (e.g., IoT nodes) \\
        
        Implementation Complexity & High & Low \\
        \bottomrule
    \end{tabular}
\end{table*}

\medskip

The constraints and simplifications of Minimal IKE enable protocol-level optimizations that reduce computational and communication overhead. A key area benefiting from these optimizations is the management of cryptographic keys over time, i.e., the \textit{key refresh mechanism}. 
In full IKEv2 deployments, key lifecycle management relies on additional exchanges: one to  generates fresh keys without repeating authentication, while the other to deletes obsolete SAs. Each rekey operation therefore adds cryptographic cost, extra messages, and often auxiliary mechanisms such as Dead Peer Detection (DPD) to maintain liveness. Minimal IKE, which does not support these exchanges, replaces in-place rekeying with a reauthentication-based mechanism (Reauth), re-executing the full protocol at fixed intervals to refresh keys, remove old SAs via \texttt{INITIAL\_CONTACT}, and reconfirm peer authenticity, while eliminating the overhead of additional exchanges and state management.
\medskip

\section{Post-quantum cryptography in Minimal IKE}
\label{sec:PQCtransition}

While IKEv2 has evolved to incorporate support for post-quantum algorithms via multiple extensions and RFCs, the Minimal IKE specification~\cite{rfc7815}—published in 2016—precedes the NIST PQC standardization process, which officially began in 2017. As a result, RFC 7815 makes no mention of quantum-resistant primitives or post-quantum integration strategies. 

With respect to their quantum-vulnerable counterparts, post-quantum cryptographic algorithms have a major bandwidth occupation since public keys, ciphertexts and signature become larger.
As an example of comparison between post-quantum algorithms and quantum-vulnerable ones, let us consider the case of digital signatures and focus on instances that have a security level of 128 bits (in the quantum domain for post-quantum systems and in the classical domain for quantum-vulnerable systems).
In Table \ref{tab:digsig_comparison} we consider the three post-quantum digital signature algorithms standardized by NIST and compare them with two widespread quantum-vulnerable algorithms: RSA and ECDSA.
From the table we see that introducing the new algorithms in the place of their predecessors has a significant cost in terms of key size and signature size.

\begin{table}
    \centering    
    \caption{Post-quantum and quantum-vulnerable digital signature algorithms and parameters for 128-bit security}
    \label{tab:digsig_comparison}
    \begin{tabular}{l l l l l}
    \toprule
                \textbf{Algorithm} & \makecell{\textbf{Quantum}\\\textbf{Security}} & \makecell{\textbf{Private key}\\(bytes)} & \makecell{\textbf{Public key}\\ (bytes)} & \makecell{\textbf{Signature}\\ (bytes)} \\
        \midrule        
        ML-DSA      & \cmark & \num{2528} &  \num{1312}   &  \num{2420}    \\ 
        FN-DSA      & \cmark & \num{1281} & \num{897}    & \num{666}     \\ 
        SLH-DSA     & \cmark & \num{64}   & \num{32}     & \num{17088}   \\ 
        RSA-3072    & \xmark & \num{384}  & \num{384}    & \num{384}     \\ 
        ECDSA-256   & \xmark & \num{32}   & \num{64}     & \num{64}      \\ 
    \bottomrule
    \end{tabular}
\end{table}

Another comparison is reported in Table \ref{tab:kem_comparison}, for the case of key encapsulation or key exchange mechanisms with $128$-bit security, considering ML-KEM and HQC-KEM for the post-quantum case in comparison with RSA-3072 and ECDH for the quantum-vulnerable case.
Also in this case, we observe that the impact of replacing a quantum-vulnerable algorithm with a post-quantum one in terms of key and ciphertext length is significant.

\begin{table}
    \centering    
    \caption{Post-quantum and quantum-vulnerable Key Encapsulation Mechanisms and Key exchanges, for 128-bit security}
    \label{tab:kem_comparison}
    \begin{tabular}{l l l l l}
    \toprule
                \textbf{Algorithm} & \makecell{\textbf{Quantum}\\\textbf{Security}} & \makecell{\textbf{Private key}\\(bytes)} & \makecell{\textbf{Public key}\\ (bytes)} & \makecell{\textbf{Ciphertext}\\ (bytes)} \\
        \midrule        
        ML-KEM      & \cmark & \num{1632} &  \num{800}   &  \num{768}    \\ 
        HQC-KEM      & \cmark & \num{2305} & \num{2249}    & \num{4433}     \\ 
        RSA-3072     & \xmark & \num{384}   & \num{384}     & \num{384}   \\ 
%        DH    & \xmark & \num{384}  & \num{384}    & \num{384}     \\ 
        X25519   & \xmark & \num{32}   & \num{32}     & \num{32}      \\ 
    \bottomrule
    \end{tabular}
\end{table}

Asymmetric cryptographic algorithms in IKEv2 are used in two phases: during \texttt{INIT} for key establishment, and during \texttt{AUTH} for peer authentication via digital signatures or certificates. It is important to note that the use of post-quantum cryptographic primitives affects the two phases of the IKEv2 handshake in very different ways. 
In fact, the \texttt{AUTH} exchange is naturally more tolerant of large cryptographic material. Since this phase supports message fragmentation, it is possible to transmit larger post-quantum signatures or certificates without modifying the protocol’s structure or exposing it to attacks. Moreover, when raw public keys are used, they can be securely distributed out-of-band. 
In contrast, the \texttt{INIT} exchange imposes much stricter constraints. Fragmentation is explicitly disallowed during this phase, primarily to prevent unauthenticated peers from sending oversized messages that could lead to denial-of-service attacks based on buffer exhaustion. 
This is particularly problematic given that post-quantum key encapsulation mechanisms (KEMs) tend to produce significantly larger messages than traditional Diffie–Hellman exchanges, often exceeding typical network MTUs. As a result, integrating post-quantum key establishment into the initial phase of IKEv2 is considerably more complex and may require structural changes to the protocol.

Standard IKEv2 addresses these issues by introducing multiple extensions, such as \texttt{RFC 8784} \cite{rfc8784}, \texttt{RFC 9242} \cite{rfc9242} and \texttt{RFC 9370} \cite{rfc9370}. While effective, these extensions increase protocol complexity and contradict the minimalist design philosophy of Minimal IKE.

For introducing post-quantum cryptography in Minimal IKE, we follow a different approach.
We extend Minimal IKE with post-quantum capabilities by integrating ML-KEM \cite{MLKEM} in the \texttt{INIT} phase.
This choice is motivated by the fact that ML-KEM is the first post-quantum key encapsulation mechanism standardized by NIST, and it is the only KEM among those standardized that can be integrated directly into the \texttt{INIT} phase without causing IP fragmentation, which would otherwise compromise the security of the protocol.
Since this exchange traditionally relies on a key agreement mechanism, adapting it to use a KEM required modifications to both the protocol semantics and payload structure. Semantically, the meaning of the \texttt{KE} payload changes — for the responder, it no longer carries a public key but rather the ciphertext produced by the encapsulation process. This ciphertext securely conveys the shared secret from which all subsequent cryptographic material for the protocol is derived. Syntactically, the payload format must also be updated to communicate the use of the new key establishment method. Further details on the integration of ML-KEM are available in \cite{ietf-ipsecme-ikev2-mlkem-00}. For the \texttt{AUTH} phase, instead, quantum-resistance can also be achieved without introducing post-quantum asymmetric cryptographic primitives, but by relying on the quantum-resistance of symmetric primitives.
To aim for the smallest possible footprint, we decided to take advantage of this opportunity, implementing authentication based on pre-shared keys (PSK) combined with HMAC \cite{HMAC}, which provides strong authenticity guarantees while preserving the minimality and simplicity that characterize Minimal IKE.
Moreover, authentication based on PSK and HMAC represents the mandatory authentication method in Minimal IKE, with all other methods being optional.
However, the use of PSK and HMAC alone is not sufficient for secure key derivation between the two endpoints. Since all the key material is long-term and shared, the compromise of a single terminal would reveal everything needed to derive the session keys and break all past and future communications. For this reason, an ephemeral Key Encapsulation Mechanism (KEM) is required to introduce fresh and ephemeral key material into the protocol, ensuring forward secrecy and limiting the impact of long-term key compromise.

By using authentication based on PSK and HMAC along with a post-quantum KEM, we can achieve resistance against an attacker model equipped with a cryptographically relevant quantum computer (CRQC), achieving NIST security level 1.

\section{Implementation and validation}
\label{sec:implementation}

Given the current lack of efficient and lightweight implementations of Minimal IKE, we developed a custom implementation from scratch, which we called \textit{Colibri}.
Implementing Colibri required addressing several protocol-specific and architectural considerations. A key issue is byte ordering: as a network protocol, all multi-byte fields must follow network byte order (big endian) according to internet standards. However, most modern devices use little endian architectures, such as x86. This necessitates careful handling of the difference between the internal representation of protocol data structures and their serialized form transmitted over the network. To address this, Colibri adopts a dual-representation approach, distinguishing between two \textit{domains}:

\begin{itemize}
    \item \textit{Logical structs}: High-level C structures used for internal processing. These are designed to facilitate data manipulation, validation, and cryptographic operations. They are independent of endianness and unconstrained by memory layout. 
    %Deals with semantic of the protocol

    \item \textit{Binary structs}: Low-level, packed representations that exactly match the IKEv2 payload format as specified in the standard. These are used exclusively for parsing incoming messages and serializing outgoing data, ensuring precise control over field alignment, size, and byte order.
    %Deals with syntax of the protocol
\end{itemize}

This separation between logical and binary representations enables a modular and layered architecture that cleanly distinguishes internal protocol logic from its wire-level encoding. The implementation is organized into independent modules, where each component is responsible for a specific functionality. Modules implementing the logical representation manage protocol semantics, including message construction, cryptographic operations, and state handling, without being constrained by the underlying byte-level format. In addition, a dedicated translation layer is responsible for converting between logical structures and their corresponding binary payloads, as specified by the IKEv2 protocol payload.

To support the semantic modeling of protocol participants and their roles in the IKE key exchange, Colibri defines a set of core logical structures, which are described in Table \ref{tab:logical-structs}. 

\begin{table}[htbp]
    \centering    
    \caption{Colibri logical structures}
    \label{tab:logical-structs}
    \begin{tabular}{@{}p{2.5cm} p{5.5cm}@{}}
        \toprule
        \textbf{Struct name} & \textbf{Description} \\
        \midrule
        \texttt{crypto\_ctx\_t}  & Captures all the cryptographic information associated with a peer. \\
        \texttt{net\_endpoint\_t} & Represents the network identity of a participant, encapsulating its file descriptor and address information. \\
        \texttt{auth\_ctx\_t} & Contains the assertion of his identity and the method that will use to prove his identity.. \\
        \bottomrule
    \end{tabular}
\end{table}

To meet the objectives of minimal memory usage and low computational overhead, the entire protocol logic and payload handling, as specified in the RFC, were implemented from scratch. However, for standard and well-established functionalities, external libraries were employed to avoid unnecessary complexity and ensure robustness. 

\subsection{Test and validation}

As explained in Section~\ref{sec:minimalIKE}, Minimal IKE supports only the Initiator role. 
Thus, in order to verify compliance with the RFC specifications, it is necessary to test its implementation by establishing communication with a fully functional Responder. To achieve this, we developed a dedicated testing environment, in which the Colibri Initiator interacts with a Dockerized instance of StrongSwan, which is a widely adopted and reference-compliant implementation of IKEv2\footnote{StrongSwan is an open source software providing fully fledged implementations of IKEv2 initators and responders \url{https://strongswan.org/}. StrongSwan is largely recognized as the most widespread, general purpose implementation of IKEv2, since it can be easily tuned (essentially, by modifying a configuration file) to simulate several cryptographic suites and configurations for IKEv2.}. This environment allows us to validate the correctness of message formatting, cryptographic operations, and overall protocol behavior against a de facto standard.

By exploiting the developed testing environment, which has been made publicly available in the Colibri repository, we are able to perform some benchmarking of Colibri to verify the architectural advantages discussed above.
For this purpose, we compare it with a StrongSwan configuration implementing a fairly generic IKEv2 Initiator.
For performing a fair comparison, in both Colibri and StrongSwan we employed the same cipher suite: SHA-1 as the pseudorandom function and authentication algorithm, AES-128 for symmetric encryption, and X25519 ECDH for key exchange in the classic variant.
For the post-quantum variant, X25519 was replaced with ML-KEM-512, with all primitives sourced from the OpenSSL library to ensure a consistent and fair baseline.
Authentication was performed using pre-shared keys (PSK). The StrongSwan configuration was minimized by disabling non-core payloads and loading only the essential protocol modules, aligning it as closely as possible with the Minimal IKE feature set.
As performance indicators we consider:

\begin{itemize}
    \item The time required to generate the \textit{Init} request.
    \item The time required to generate the \textit{Auth} request.
    \item The overall memory occupied during both the \textit{Init} and \textit{Auth} phases.
\end{itemize}

As far as execution times are concerned, it should be noted that they are also influenced by the number of threads used by each software. In fact, while StrongSwan is multi-threaded software that requires at least 6 available threads to run, Colibri was written as a single-threaded program. 
Clearly, this makes the comparison not entirely fair; however, it represents the only possible basis for comparison, since it is not possible to obtain a single-threaded instance of StrongSwan.
In fact, this difference is intrinsic to the design goals of the two implementations: StrongSwan aims to provide the widest possible support for all the functionalities offered by the IKEv2 protocol, whereas Colibri is designed to implement only the strictly necessary features included in Minimal IKE, doing so in the lightest and most streamlined way possible.
Therefore, the reported results should be interpreted with this consideration in mind.

For the case using classic (quantum-vulnerable) cryptographic primitives, Table \ref{tab:benchmark_classic} reports the size in bytes and the corresponding generation time for each request, in addition to the memory resources required to complete the exchange.
As we can see from the table, for the Init phase, Colibri requires a shorter request packet, while for the authentication phase the two request packets have the same length.
As regards the generation time for such requests, we should take into account that StrongSwan requires at least 6 threads to run, while Colibri uses a single thread.
Despite this, the time taken by Colibri is comparable to, if not less than, that taken by StrongSwan.  
The main advantage of Colibri over StrongSwan results from its memory usage, which is reduced by about 5 and 4 times compared to StrongSwan, in terms of average and peak values, respectively.

\renewcommand{\arraystretch}{1.3}
\begin{table}[htbp]
    \centering    
    \caption{Comparison of Minimal IKE and IKEv2 performance\\ (classical cipher suite)}
    \label{tab:impl-characteristics}
    \begin{tabularx}{0.48\textwidth}{l l l l}
        \toprule
        \textbf{Phase} & \textbf{Metric} & \textbf{Colibri} & \textbf{StrongSwan} \\
        \midrule
         & No. threads & 1 & 6 \\
         \midrule
        \multirow{2}{*}{\textit{Init Req.}} 
        & Bytes    & \num{152} B       & \num{240} B \\
        & Generation Time & \num{1.19} ms    & \num{0.954} ms \\
        \addlinespace[0.5em]

        \multirow{2}{*}{\textit{Auth Req.}} 
        & Bytes    & \num{108} B       & \num{108} B \\
        & Generation Time & \num{0.364} ms    & \num{0.448} ms \\
        \addlinespace[0.5em]

        \multirow{2}{*}{\textit{Overall}} 
        & Avg Mem Usage  & $1.28 \pm 0.27$ MB       & $6.21 \pm 0.64 $ MB \\
        & Peak Memory   & $1.67 \pm 0.37$  MB     & $6.62 \pm 0.69$ MB \\
        \bottomrule
    \end{tabularx}
    \label{tab:benchmark_classic}
\end{table}

For the case considering post-quantum cryptographic primitives, the same performance indicators have been assessed and are reported in Table \ref{tab:benchmark_pqc}.
From the table, we can see that, as expected, the transition to post-quantum cryptography yields an increase in the size of messages to be exchanged and the time required to generate them.
In this case, StrongSwan requires longer messages to perform both the Init and Auth requests than Colibri.
This is also reflected in the timing, which is longer for StrongSwan than for Colibri, despite the fact that StrongSwan uses 6 threads compared to the single thread of Colibri.
Finally, also in this case, Colibri exhibits a significant advantage over StronSwan in terms of memory usage, which is between 3 and 4 times lower.

\begin{table}[htbp]
    \centering    
    \caption{Comparison of Minimal IKE and IKEv2 performance\\ (post-quantum cipher suite)}
    \label{tab:impl-characteristics-pqc}
     \begin{tabularx}{0.48\textwidth}{l l X X}
        \toprule
        \textbf{Phase} & \textbf{Metric} & \textbf{Colibri} & \textbf{StrongSwan} \\
        \midrule
        & No. threads & 1 & 6 \\
        \midrule

        \multirow{2}{*}{\textit{Init Req.}} 
        & Bytes    & \num{920} B       & \num{1164} B \\
        & Generation Time & \num{1.116} ms    & \num{1.764} ms \\
        \addlinespace[0.5em]

        \multirow{2}{*}{\textit{Auth Req.}} 
        & Bytes    & \num{108} B       & \num{172} B \\
        & Duration & \num{0.352} ms    & \num{0.493} ms \\
        \addlinespace[0.5em]

        \multirow{3}{*}{\textit{Overall}} 
        & Avg Mem Usage  & $1.75 \pm 0.35$ MB         &  $6.67 \pm 0.37 $ MB  \\
        & Memory Peak   & $2.13 \pm 0.42$ MB              &  $6.99 \pm 0.40 $ MB \\
        \bottomrule
    \end{tabularx}
    \label{tab:benchmark_pqc}
\end{table}

\section{Conclusion and future work}
\label{sec:conclusion}

We considered the Minimal IKE protocol, which had not been efficiently implemented since its introduction, and revitalized it by both providing an efficient, open-source implementation and introducing a post-quantum variant of it. 
The tools we developed, which we made publicly available, allowed us to verify the advantages in terms of complexity that this protocol can have over the fully fledged IKEv2 protocol, allowing it to be adopted even in contexts characterized by limited resources, and without sacrificing quantum resistance.
Among the application scenarios of interest are all networks in which devices with reduced computing capabilities must be used, including modern satellite networks, which use security protocols similar to those of terrestrial networks, but with obviously greater resource limitations \cite{GSM2026}.

As a direction for future work, we plan to include further options for transitioning Minimal IKE to post-quantum cryptography while preserving its simplicity and minimal message flow, including the hybrid use of classical and post-quantum cryptography for a transitional period.
Regarding authentication, we plan to extend the AUTH mechanism to support KEM-based authentication instead of traditional digital signatures. 

\printbibliography

\end{document}